\documentclass[aps,prl,twocolumn,superscriptaddress]{revtex4-2}

\usepackage{graphicx}
\usepackage[dvips]{color}
\usepackage{braket}
\usepackage{color} 
\usepackage{ulem}

\begin{document}


\title{Robust Metal-Insulator Transition Despite Surface Dead-Layer Growth in Sub-10-nm Cr-Doped V$_2$O$_3$ Nanocrystals}


\author{Yoichi Ishiwata}
\email{ishiwata@cc.saga-u.ac.jp}
\author{Ichidai Harada}
\affiliation{Department of Physics, Saga University, Saga 840-8502, Japan}
\author{Masaki Imamura}
\author{Kazutoshi Takahashi}
\affiliation{Synchrotron Light Application Center, Saga University, Saga 840-8502, Japan}
\author{Hirofumi Ishii}
\author{Masato Yoshimura}
\author{Nozomu Hiraoka}
\affiliation{National Synchrotron Radiation Research Center, Hsinchu 30076, Taiwan}
\author{Yuji Inagaki}
\affiliation{Institute for the Advancement of Higher Education, Okayama University of Science, Okayama 700-0005, Japan}
\author{Kenta Akashi}
\author{Tatsuya Kawae}
\affiliation{Department of Applied Quantum Physics, Kyushu University, Fukuoka 819-0395, Japan}
\author{Tetsuya Kida}
\affiliation{Department of Applied Chemistry and Biochemistry, Kumamoto University, Kumamoto 860-8555, Japan}
\author{Masashi Nantoh}
\affiliation{Pioneering Research Institute, The Institute of Physical and Chemical Research (RIKEN), Wako, 351-0198, Japan}
\affiliation{RIKEN Nishina Center for Accelerator-Based Science (RNC), Wako, 351-0198, Japan}



\date{\today}

\begin{abstract}

We investigated the size dependence of the metal-insulator transition (MIT) in Cr-doped V$_2$O$_3$ nanocrystals by photoemission spectroscopy using complementary probing depths, together with magnetic susceptibility measurements. Photoemission spectra show that MIT signatures persist down to an average particle size of 5.6 nm, and magnetic susceptibility measurements exhibit a nearly size-invariant transition onset. The contrast between surface-sensitive and deeper-probing photoemission spectra reveals that the transition survives in the nanocrystal interior. At the same time, the spectra indicate a systematic suppression of coherent quasiparticle weight with decreasing size, pointing to the growth of an insulating surface dead layer. These results demonstrate that nanoscaling does not intrinsically eliminate the MIT itself, but progressively enhances the influence of surface-driven insulating behavior, thereby providing insight into the practical limits of miniaturizing Mott-based devices.

\end{abstract}


\maketitle

Phase transitions in correlated electron systems have attracted broad interest because of their relevance to fundamental physics and emerging technologies \cite{Paschen}. In particular, understanding how these transitions evolve at the nanoscale is crucial for both theory and device applications \cite{Li}. The influence of particle size on phase transitions has been extensively studied in magnetism \cite{Lang, Sun}.  In magnetic nanoparticles, the transition temperature generally decreases for smaller particle sizes, a trend commonly attributed to the increasing fraction of undercoordinated surface atoms that weakens exchange interactions. Such behavior is often rationalized by cohesive-energy-based models and exemplifies a broader class of surface-driven size effects in nanoscale systems \cite{Lang, Sun}.

Whether an analogous picture holds for the metal-insulator transition (MIT) is much less clear. The MIT is rooted in electronic correlation and band filling \cite{Mott, Tsuda, IFT}, and is not determined solely by local coordination effects at the surface, although reduced coordination at the surface can itself locally enhance correlation effects. Furthermore, in many cases, the transition is  first order and involves coupled electronic, magnetic, and structural degrees of freedom \cite{Mott, Tsuda, IFT}. Its size dependence is therefore not expected to be universal and may vary from material to material depending on which degree of freedom plays the dominant role. Despite these considerations, experimental studies on the size dependence of the MIT remain scarce.

From a theoretical perspective, the intrinsic criterion for a Mott transition is set primarily by bulk electronic parameters. In the classical Mott picture, the transition boundary is governed by the carrier density and the effective Bohr radius \cite{Mott, Tsuda, IFT}. A doublon-holon picture offers a more microscopic view in which the relevant scale emerges from electronic correlation \cite{Yokoyama}. These viewpoints suggest that the MIT itself need not disappear simply because the system size is reduced. At the same time, however, strong correlation can be enhanced near a surface, where reduced coordination modifies the local electronic environment. In correlated oxides, this effect can produce an insulating surface ``dead layer'' \cite{Borghi}. Thus, in nanoscale Mott systems, size reduction raises two distinct but closely related questions: whether the transition in the particle interior remains intrinsically robust, and how the increasing surface fraction influences the overall electronic behavior.

V$_2$O$_3$ is a prototypical Mott system that exhibits a first-order MIT accompanied by antiferromagnetic and structural transitions in the bulk \cite{Mott, Tsuda, IFT}. Our previous studies on V$_2$O$_3$ nanocrystals (NCs) showed that the MIT is suppressed in undoped particles at sizes of $\sim$20 nm and above, whereas impurity doping such as Cr or Ti allows the transition to persist  \cite{Ishiwata2012APL,IshiwataAMI,Ishiwata2020PRB}. It remains unknown, however, whether the transition can still survive at particle sizes below $\sim$10 nm. Meanwhile, photoemission studies have shown that V$_2$O$_3$ possesses an insulating surface layer whose thickness reaches several nanometers \cite{Rodolakis}, consistent with the general theoretical expectation of enhanced correlation at a surface \cite{Borghi}. For sub-10 nm NCs, this length scale becomes comparable to the particle size itself. Therefore, determining whether the MIT is truly suppressed by size reduction or instead survives in the particle interior despite the growing influence of the insulating surface region is a central issue for understanding size effects in Mott NCs and for assessing the scalability of Mott-based devices \cite{Zhou, Wang, Ran}.

Here, we investigate Cr-doped V$_2$O$_3$ NCs down to 5.6 nm by photoemission spectroscopy using two complementary probing depths, together with magnetic susceptibility measurements. This approach allows us to distinguish the behavior of the NC interior from that of the near-surface region. Our results reveal that the MIT survives in the NC interior, while the insulating surface dead layer thickens with decreasing size. 

\begin{figure*}
\begin{center}
\includegraphics[scale=0.5]{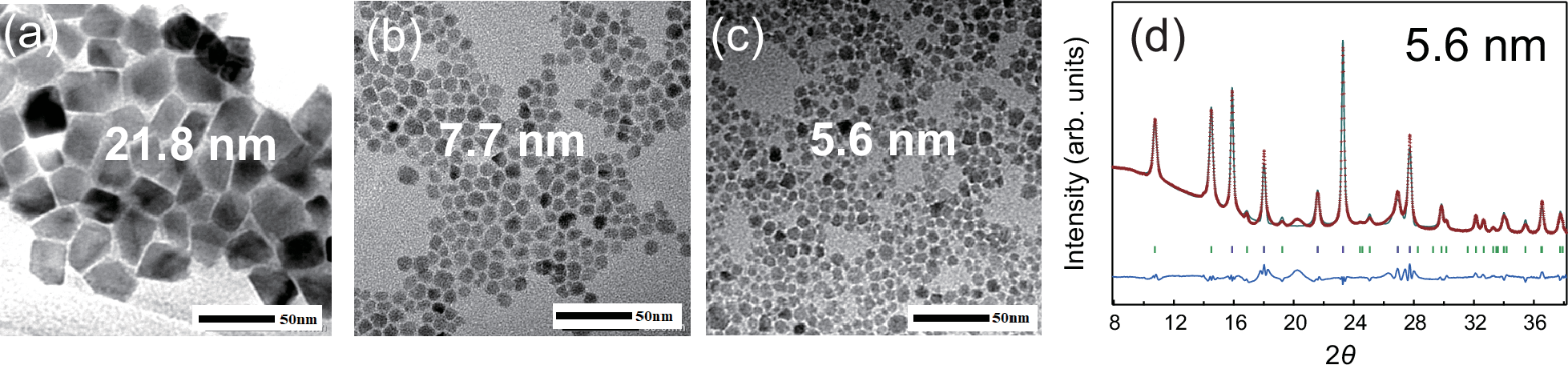}
\end{center}
\caption{TEM images for Cr-doped V$_2$O$_3$ NCs with average sizes of (a) 21.8 $\pm$ 3.8 nm, (b) 7.7 $\pm$ 2.1 nm, and (c) 5.6 $\pm$ 2.0 nm. (d) Synchrotron XRD pattern of the 5.6 nm sample with Rietveld refinement confirming the corundum structure. }
\label{f1}
\end{figure*}

Cr-doped V$_2$O$_3$ NCs were synthesized by a modified two-step procedure based on our previous work \cite{IshiwataAMI,Ishiwata2020PRB}, consisting of a sealed solvothermal treatment followed by high-temperature decomposition. The Cr content was adjusted to $\sim$1\%. NCs with average sizes of 5.6 $\pm$ 2.0 nm [Fig. 1(c)] and 7.7 $\pm$ 2.1 nm [Fig. 1(b)] were obtained under identical heating conditions, with the size difference controlled by the subsequent cooling process: the former sample was rapidly cooled, whereas the latter was allowed to cool naturally. A portion of the 5.6 nm sample was further annealed to yield NCs with an average size of 21.8 $\pm$ 3.8 nm [Fig. 1(a)].

Figures 1(a)-1(c) show transmission electron microscopy (TEM) images of NCs. The largest sample (21.8 nm) exhibits clear faceting, indicative of particle coalescence during the post-synthesis annealing. Similar facet development has been reported for V-doped ZnO  NCs and CoO NCs in our earlier studies \cite{Tsukahara, IshiwataCoO}, suggesting a common mechanism driven by surface energy minimization. Synchrotron x-ray diffraction (XRD) was performed at room temperature at BL12B2 of SPring-8 using an incident wavelength of 0.6889 \AA, and the data were analyzed by Rietveld refinement with RIETAN-FP \cite{Izumi}. The XRD pattern of the smallest sample [Fig. 1(d)] confirms that the corundum structure \cite{Mott, Tsuda, IFT} is preserved even at 5.6 nm.

DC magnetization was measured at 5000 Oe using a SQUID magnetometer [magnetic property measurement system (MPMS), Quantum Design].

Hard x-ray photoelectron spectroscopy (HAXPES) and soft x-ray photoelectron spectroscopy (SXPES) were performed at BL12XU of SPring-8 ($h\nu = 6516$ eV) and BL13 of SAGA-LS ($h\nu = 530$ eV), respectively. The probing depths were estimated to be approximately 7.8 nm for HAXPES and 1.1 nm for SXPES using inelastic mean free path values for elemental V at photoelectron kinetic energies relevant to the valence-band measurements \cite{Tanuma}. The energy resolution was about 300 meV in both cases. Measurements were made during warming after cooling to $\sim$10 K (SXPES) and $\sim$80 K (HAXPES). The measurement geometries were $(\theta; \phi) = (90^\circ; 90^\circ)$ for HAXPES and $(45^\circ; 180^\circ)$ for SXPES, where $\theta$ and $\phi$ are defined using the photon polarization, photon propagation, and photoelectron emission directions \cite{Trzhaskovskaya}. NC samples were drop-cast onto highly oriented pyrolytic graphite substrates, followed by ligand removal using a Meerwein reagent \cite{Suehiro}.

\begin{figure}
\begin{center}
\includegraphics[scale=0.8]{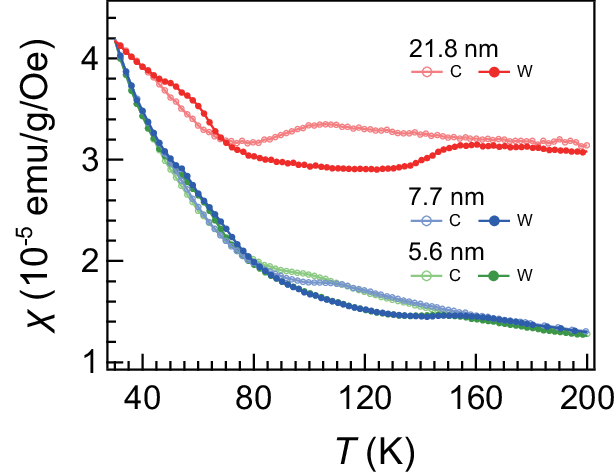}
\end{center}
\caption{Temperature-dependent magnetic susceptibility of Cr-doped V$_2$O$_3$ NCs with sizes of 5.6 nm, 7.7 nm, and 21.8 nm. Warming curves exhibit a clear increase near 155 K, indicative of the antiferromagnetic-to-paramagnetic transition.}
\label{f2}
\end{figure}

Figure 2 shows the temperature-dependent magnetic susceptibility of Cr-doped V$_2$O$_3$ NCs. Because the transition is first order, cooling and warming curves exhibit hysteresis arising from a nucleation barrier between the old and new phases \cite{Binder}. The warming curves are focused on below because thermal activation reduces the influence of this barrier and provides a better approximation to the intrinsic transition temperature. All samples show a clear increase in susceptibility near 155 K during warming, indicative of the antiferromagnetic-to-paramagnetic transition, as established in bulk V$_2$O$_3$ \cite{Mott, Tsuda, IFT}. This nearly size-invariant transition temperature contrasts with the well-known tendency for magnetic transition temperatures to decrease as the particle size is reduced \cite{Lang, Sun}, suggesting that the magnetic transition in these NCs is not independently controlled by particle size but is instead driven by another coupled transition, most likely the MIT, as supported by the photoemission results discussed below.

A hysteresis width of several tens of kelvin is typical of first-order transitions in nanoscale systems and is often associated with the scarcity of extended defects that would otherwise facilitate nucleation of the new phase \cite{Appavoo}. In high-quality NCs, the absence of such defects suppresses nucleation and can broaden the hysteresis, as previously reported for CdSe NCs \cite{Chen}, VO$_2$ nanoparticles \cite{Appavoo} or V$_2$O$_3$ NCs \cite{IshiwataAMI,Ishiwata2020PRB}. Cr doping also promotes the transition, in contrast to undoped V$_2$O$_3$ NCs of comparable size, where it is completely suppressed \cite{IshiwataAMI}. The susceptibility of the 21.8 nm sample is unexpectedly higher than that of the smaller NCs, possibly reflecting structural evolution such as facet formation. A minor increase in susceptibility between 45 and 80 K is also observed; while residual oxygen in the SQUID chamber may contribute, an intrinsic origin cannot be ruled out.

\begin{figure}
\begin{center}
\includegraphics[scale=0.45]{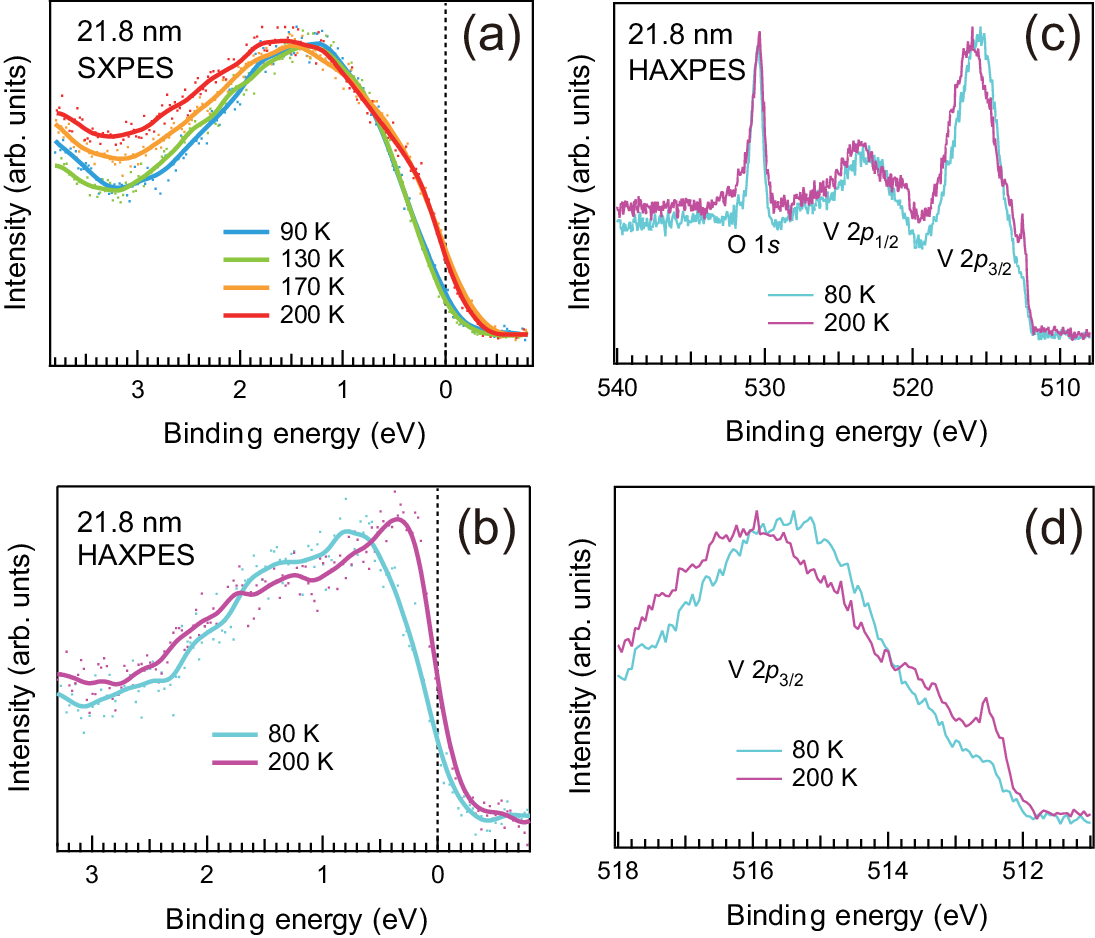}
\end{center}
\caption{Photoemission spectra of the 21.8 nm sample. (a) SXPES valence-band spectra showing temperature-induced changes, including the emergence of a shoulder feature near $E_{\rm{F}}$ above 170 K. (b) HAXPES spectra revealing a clear metallic response at 200 K. (c,d) V 2$p$ and O 1$s$ core-level spectra, which exhibit temperature-dependent changes consistent with the MIT.}
\label{f3}
\end{figure}

Figure 3 summarizes the photoemission results for the 21.8 nm sample. Figure 3(a) shows surface-sensitive SXPES valence-band spectra.  At 90 and 130 K, the intensity near the Fermi level ($E_{\rm{F}}$) is strongly suppressed, indicating an insulating phase with an opened gap. At 170 and 200 K, the spectral weight at $E_{\rm{F}}$ increases and a shoulder feature associated with the quasiparticle (QP) emerges, signaling the metallic phase \cite{Mo, Panaccione, Papalazarou}. 

Figure 3(b) shows deeper-probing HAXPES valence-band spectra at 80 and 200 K. The 80 K spectrum is insulating-like but retains slight intensity near $E_{\rm F}$, possibly because the gap does not fully open at this temperature. At 200 K, a pronounced QP peak appears, consistent with bulk V$_2$O$_3$ \cite{Mo, Panaccione, Papalazarou}. Compared with SXPES, the deeper probing in HAXPES enhances the metallic spectral weight from the NC interior, supporting the presence of an insulating surface layer (dead layer), consistent with previous photoemission studies on bulk V$_2$O$_3$ \cite{Rodolakis, Mo}.

Figure 3(c) shows the wide-range core-level spectrum including the V 2$p$ and O 1$s$  regions, while Fig. 3(d) enlarges the V 2$p_{3/2}$ region. In Fig. 3(c), shakedown satellites appear at 200 K in both the V 2$p_{3/2}$ and V 2$p_{1/2}$ regions, near 512.5 and 520.5 eV, respectively. These satellites originate from efficient screening of the core hole by itinerant electrons in the metallic phase  \cite{Panaccione} and disappear at 80 K, where such screening is strongly suppressed. These core-level changes indicate that itinerant-electron screening is already strongly suppressed at 80 K, even though the valence-band spectrum still retains slight intensity near $E_{\rm F}$.

\begin{figure}
\begin{center}
\includegraphics[scale=0.45]{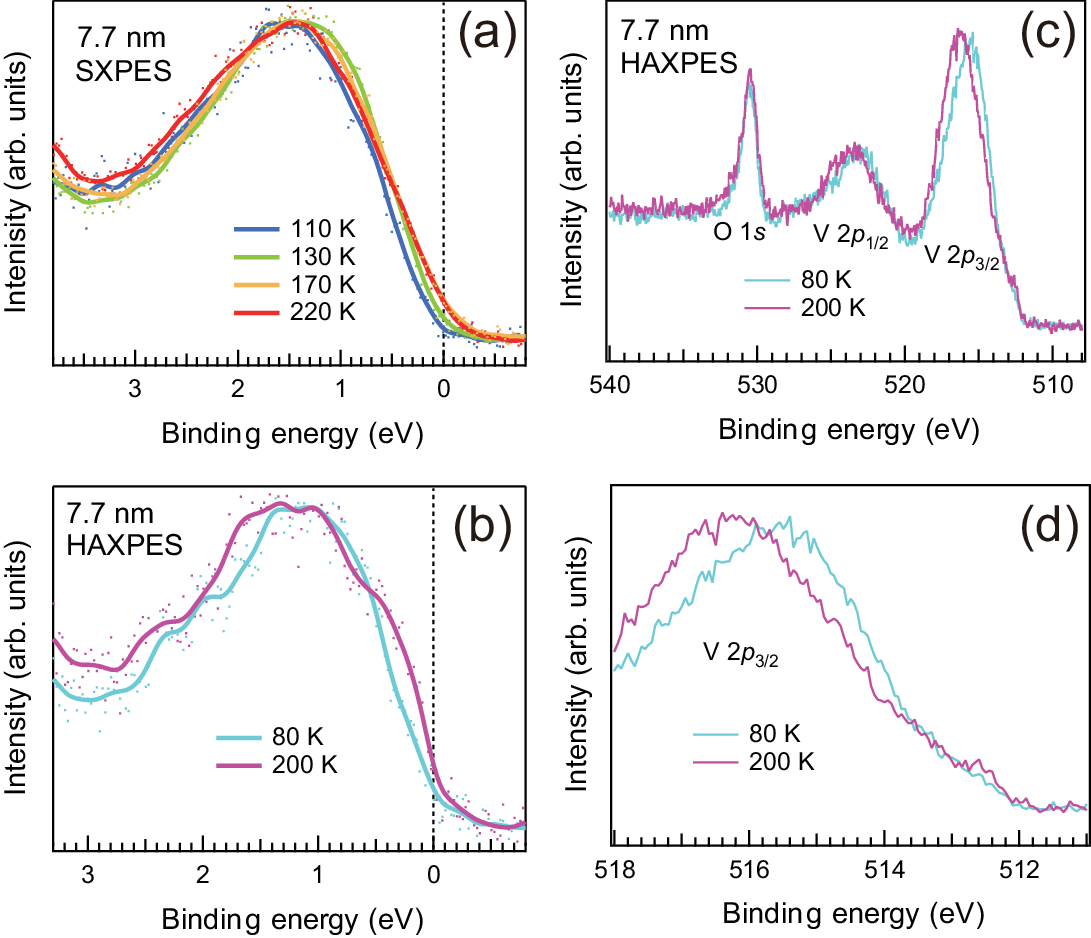}
\end{center}
\caption{Photoemission spectra of the 7.7 nm sample. (a) SXPES valence-band spectra showing minimal temperature-induced changes across the measured range. (b) HAXPES spectra reveal a metallic response at 200 K despite strong surface suppression. (c,d) V 2$p$ and O 1$s$ core-level spectra exhibit small but discernible changes between 80 and 200 K, consistent with a weakened MIT signature.}
\label{f4}
\end{figure}

Figures 4 and 5 summarize the photoemission results for the 7.7 and 5.6 nm samples, respectively. In both cases, SXPES valence-band spectra exhibit strongly suppressed intensity near $E_{\rm F}$ with only minimal temperature-dependent change, indicating that the near-surface region remains predominantly insulating. The suppression is stronger in the 5.6 nm sample, where no clear QP shoulder is observed even at 220 K, suggesting that the region probed by SXPES lies entirely within the insulating surface layer.

The deeper-probing HAXPES spectra nonetheless retain weak MIT signatures in both sub-10 nm samples. For 7.7 nm NCs, the 200 K spectrum exhibits a shoulder near $E_{\rm F}$ without developing a distinct QP peak, indicating that the NC interior still undergoes a MIT despite strong surface suppression. For 5.6 nm NCs, the corresponding spectral change is even smaller, but a finite increase in spectral weight near $E_{\rm F}$ remains detectable at 200 K, implying that an electronically active core still survives at this extreme size.

The core-level spectra are consistent with this trend. In the 7.7 nm sample, the V 2$p_{3/2}$ region exhibits a small temperature-dependent change between 80 and 200 K, whereas in the 5.6 nm sample the corresponding change becomes extremely weak. Together, these results show that MIT signatures persist below 10 nm, while the insulating surface contribution grows rapidly with decreasing size. 

\begin{figure}
\begin{center}
\includegraphics[scale=0.45]{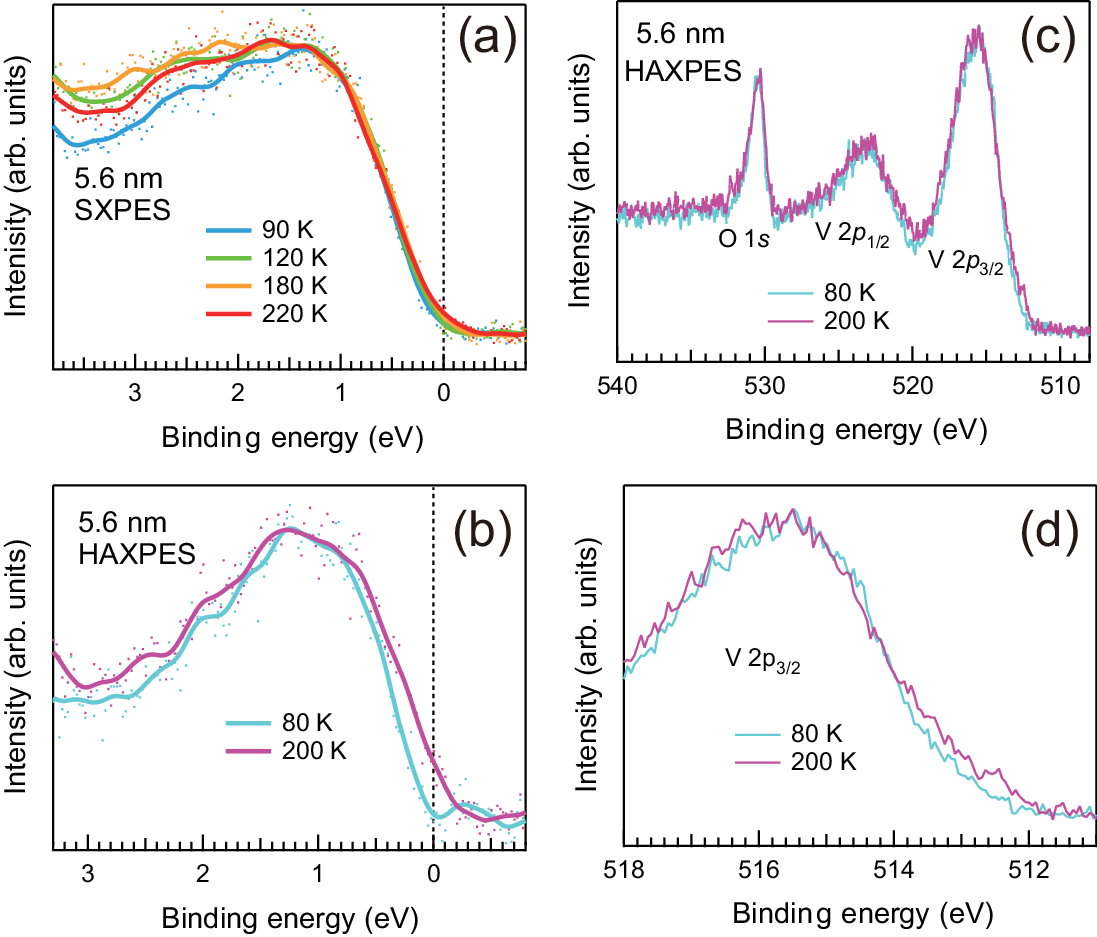}
\end{center}
\caption{Photoemission spectra of the 5.6 nm sample. (a) SXPES valence-band spectra remain insulating-like across all temperatures, indicating that the probed region is entirely within the insulating surface layer. (b) HAXPES spectra show a slight increase in spectral weight near $E_{\rm F}$ at 200 K, suggesting that the NC interior undergoes a MIT even at this extreme size. (c,d) V 2$p$ and O 1$s$ core-level spectra exhibit subtle changes between 80 and 200 K, confirming the persistence of the MIT internally.}
\label{f5}
\end{figure}

\begin{figure}
\begin{center}
\includegraphics[scale=0.25]{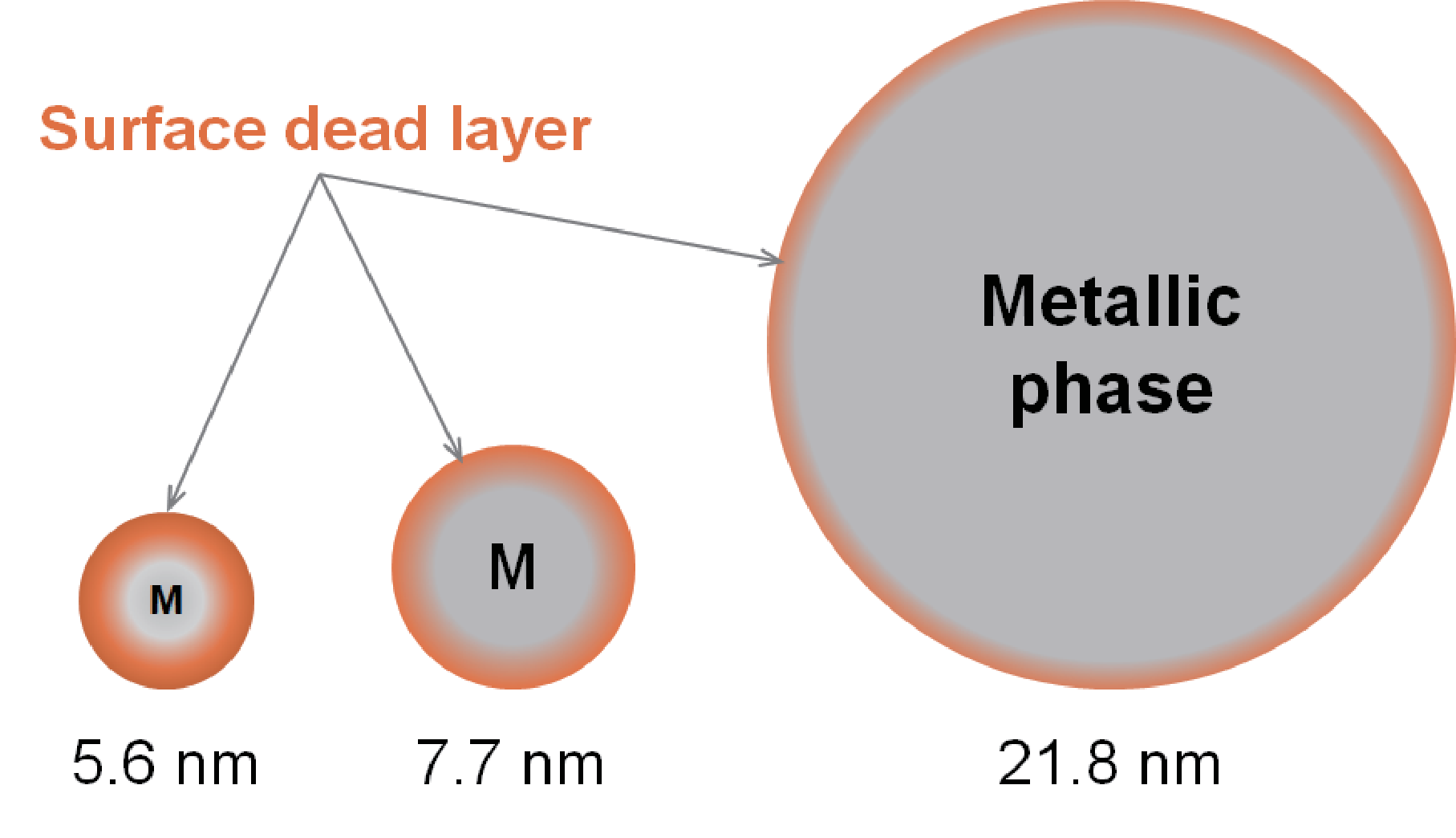}
\end{center}
\caption{Schematic illustration of the two-layer model: an insulating surface layer surrounds a metallic interior in Cr-doped V$_2$O$_3$ NCs. As size decreases, the insulating layer thickens and strongly influences both surface-sensitive and deeper-probing photoemission spectra.}
\label{f6}
\end{figure}

Our results demonstrate that reducing the particle size in Cr-doped V$_2$O$_3$ NCs does not eliminate the MIT, but progressively enhances the influence of the insulating surface region on the photoemission response. The 5.6 nm sample provides the clearest example: surface-sensitive SXPES remains insulating-like over the entire temperature range, whereas the more deeply probing HAXPES still exhibits a finite increase in spectral weight near $E_{\rm F}$ at 200 K. Together with the size-invariant magnetic transition near 155 K (Fig. 2), these results indicate that the intrinsic transition survives even at this scale, although its observable spectral signature is increasingly masked by the insulating surface region.

The nearly size-invariant transition temperature is in stark contrast to the strong size dependence commonly observed for magnetic transitions \cite{Lang, Sun}, but is consistent with the view that the transition is driven primarily by electronic correlation within the NC interior. According to the classical Mott criterion, the critical boundary satisfies $n^{1/3} a_{B} \approx 0.25$ \cite{Mott, Tsuda, IFT}, where $n$ is the carrier density and $a_{B}$ is the effective Bohr radius of the carriers. This shows that the occurrence of the MIT is determined by the average carrier spacing and the characteristic spatial extent of the carrier wave function, the latter being closely related to electronic correlation. A doublon-holon picture emphasizes that the transition occurs when the doublon-holon binding length becomes comparable to the minimum doublon-doublon (holon-holon) separation, both set by the strength of electronic correlation \cite{Yokoyama}. Within these frameworks, simple size reduction does not necessarily eliminate the MIT itself. 

However, at the same time, reduced coordination and weakened screening at the surface strengthen electronic correlation in the near-surface region, and promote the formation of an insulating surface layer (dead layer) \cite{Borghi, Rodolakis}, an intrinsic consequence of having a surface in strongly correlated systems such as V$_2$O$_3$ and therefore an unavoidable feature. In the 21.8 nm sample, the SXPES valence-band spectrum shows a QP feature only as a shoulder near $E_{\rm F}$, whereas bulk V$_2$O$_3$ has a clear QP peak at similar photon energies \cite{Mo}. This contrast indicates that even at 21.8 nm the insulating surface layer is thicker than in bulk, as evidenced by the known excitation-energy dependence in bulk V$_2$O$_3$, where the QP reduces to a shoulder only at much lower photon energies (e.g., 60 eV) \cite{Mo}. The progressive weakening of the QP feature in the 7.7 nm and 5.6 nm samples further indicates that the insulating surface layer becomes thicker as particle size decreases. 

One possible explanation for this thickening is suggested by the surface-correlation picture of Borghi \textit{et al.} \cite{Borghi}. In that work, increasing the local electron-electron repulsion $U$ at the surface to a value corresponding to the insulating state suppresses the QP not only exactly at the surface but also over a finite depth below it, implying that surface-enhanced correlation propagates into the interior over a characteristic length scale. In NCs, where the surfaces acquire finite curvature, an interior point can be influenced by a broader surrounding portion of the surface than in a flat geometry. This geometrical effect would naturally extend the region over which the QP is suppressed, leading to an effective thickening of the dead layer. In addition, the increased curvature of small NCs may increase the density of step- and edge-like sites associated with stronger undercoordination, which could further enhance the local surface $U$ and reinforce the dead-layer expansion.

The 5.6 nm NCs further allow an estimate of the characteristic sizes of both the insulating surface dead layer and the electronically active core. Because the probing depth of SXPES is about 1.1 nm \cite{Tanuma} and the corresponding spectra remain insulating-like even at 220 K, the dead layer must extend at least across the near-surface region sampled by SXPES. At the same time, the observation of a weak metallic response in HAXPES at 200 K indicates that an electronically active core still remains. Taking into account the particle-size distribution of 5.6 $\pm$ 2.0 nm, these observations suggest that the dead-layer thickness is on the scale of a few nanometers. This, in turn, indicates that a core region on the scale of a few nanometers is sufficient to sustain the MIT, although the transition may remain operative even for smaller cores. Figure 6 summarizes this interpretation by a two-layer model consisting of an insulating surface layer and a metallic core that undergoes the MIT.

Although a dead-layer thickness of about 4 nm was previously inferred for bulk V$_2$O$_3$ from angular-dependent photoemission \cite{Rodolakis}, that value likely overestimates the effective thickness relevant to the present NCs. This may be related to the strong sensitivity of QP suppression to local surface morphology in bulk photoemission spectra \cite{Mo}. 

In summary, photoemission and magnetic susceptibility measurements on Cr-doped V$_2$O$_3$ NCs reveal that the MIT persists down to an average particle size of 5.6 nm, even though the insulating surface dead layer becomes thicker with decreasing size. These results show that nanoscaling does not intrinsically eliminate the MIT itself, but instead progressively restricts it to a smaller active core as the surface dead layer grows. The present results suggest that the fundamental particle size limit for sustaining the MIT is around 5 nm, whereas the practical limit for electronic device operation is likely larger because the electronically active core in this size range may become too small to yield a sufficiently large conductance change.

\bigskip

$Acknowledgments$---We would like to thank H. Kurita for her assistance in the early stage of this study. The XRD and HAXPES experiments at SPring-8  were conducted with the approval of JASRI (Proposal No. 2022B4257, No. 2024A4130,  No. 2024B4137, and No. 2025A4140). The SXPES experiments were performed with the approval of SL-Center, Saga University (Proposal No. R1-303V, No. R2-204V, and No. R2-303V). TEM analyses were carried out using JEM-2100 microscopes at the Analytical Research Center for Experimental Sciences, Saga University. We would like to thank Y. Tokuyama for assistance with TEM measurements. This work was supported by JSPS KAKENHI Grant Number JP25K08419, and by Institute of Industrial Nanomaterials, Kumamoto University.


\end{document}